\documentclass[default]{sn-jnl}

\usepackage{graphicx}%
\usepackage{multirow}%
\usepackage{amsmath,amssymb,amsfonts}%
\usepackage{amsthm}%
\usepackage{mathrsfs}%
\usepackage[title]{appendix}%
\usepackage{xcolor}%
\usepackage{textcomp}%
\usepackage{manyfoot}%
\usepackage{booktabs}%
\usepackage{caption}
\usepackage{cleveref}
\usepackage{braket} 
\usepackage{algorithm}%
\usepackage{algorithmicx}%
\usepackage{algpseudocode}%
\usepackage{listings}%

\graphicspath{{figures/}}
\usepackage{hyphenat}
\usepackage{siunitx}

\usepackage{soul, color, xcolor}

\theoremstyle{thmstyleone}%
\theoremstyle{thmstyletwo}%

\theoremstyle{thmstylethree}%

\begin{document}

\title[Article Title]{Security analysis of orthogonal state attack on a high-speed quantum key distribution system}



\author[2,6]{Jialei Su}
\equalcont{These authors contributed equally to this work.}

\author[2]{Qingquan Peng}
\equalcont{These authors contributed equally to this work.}

\author[3,4]{Jia-Lin Chen}
\equalcont{These authors contributed equally to this work.}

\author[3,4]{Feng-Yu Lu}

\author[2]{Zihao Chen}

\author[2]{Junxuan Liu}

\author[3,4,5]{De-Yong He}

\author*[3,4,5]{Shuang Wang}\email{wshuang@ustc.edu.cn}

\author*[1,2]{Anqi Huang}\email{angelhuang.hn@gmail.com}

\affil[1]{College of Electronic Science and Technology, National University of Defense Technology, Changsha 410073, People's Republic of China}

\affil[2]{College of Computer Science and Technology, National University of Defense Technology, Changsha 410073, People's Republic of China}

\affil[3]{Laboratory of Quantum Information, University of Science and Technology of China, Hefei 230026, China}

\affil[4]{Anhui Province Key Laboratory of Quantum Network, University of Science and Technology of China, Hefei 230026, China}

\affil[5]{Hefei National Laboratory, University of Science and Technology of China, Hefei 230088, China}

\affil[6]{School of Automation, Central South University, Changsha 410083, China}


\abstract{High-speed quantum key distribution (QKD) systems have achieved repetition frequencies above gigahertz through advanced technologies and devices, laying an important foundation for the deployment of high-key-rate QKD system. 
Although these advanced systems may introduce potential loopholes, an eavesdropper Eve is challenging to exploit them by performing the intercept-resend attacks due to the limited time window under high repetition frequency. However, here, we propose a security analysis model of orthogonal state attacks that do not require intercept-resend operation on the key rate of a QKD system. Under this framework, we propose a muted attack and experimentally verify the feasibility of the attack using a \SI{1}{\giga\hertz} single-photon avalanche detector (SPAD). By sending hundreds of photons each time, Eve can mute Bob's SPADs to control the overall detection response of the QKD receiver, allowing her to learn nearly all the keys. Furthermore, we use this security model to simulate the overestimated key rates of the QKD system under orthogonal state attacks, including both the muted attack and the dead-time attack. This work theoretically and experimentally shows a timely case of the security vulnerability in the high-speed QKD system.

}

\maketitle

\section*{Introduction}\label{sec1}
Quantum key distribution (QKD), based on the laws of quantum physics, distributes a secret random bit string between two separate parties (Alice and Bob) and provides information-theoretic security~\cite{1991bell,anquan1,anquan2,anquan3,2014BENNETT}.
In the development of QKD technology, enhancing the secret key rate remains one of the paramount focuses~\cite{Diamanti2016,Sasaki2017,Chen2021}, which stimulates various technological approaches to improve the efficiency of the system~\cite{NURUL2017,Yuan:18,BOARON2018,Li2023,DU2023,Grünenfelder2023}.
So far, QKD systems achieving repetition frequency above gigahertz are largely based on the prepare-and-measure QKD protocols~\cite{Dixon2008,Sibson2017,Wang2018,An2019,Li2023} with technological advancements in single-photon detection~\cite{Namekata06,walenta2012,YUAN2007,HE2017,he2023}. For example, the single photon avalanche detector~(SPAD) that employs the low-pass filter (LPF) and the width discriminator allows the repetition frequency to reach~\SI{2.5}{\giga\hertz}~\cite{he2023}.

Regarding the security of high-speed QKD systems, it is challenging for an eavesdropper to conduct an active attack due to the limited time window to perform intercept-resend operations under the high repetition frequency. However, if Eve can conduct an attack in which no intercept-resend operation is required, this issue would be a high risk to the security of a high-speed QKD system. Until now, only one active attack, named dead-time attack~\cite{deadtimeattack2011}, is known as a non-intercept-resend attack on the receiver side, which fortunately is relatively easy to defend against. Nevertheless, today, advanced technologies developed in the high-speed QKD system may introduce potential security vulnerabilities that make the non-intercept-resend attack possible again. More importantly, the security model considering non-intercept-resend attack on a prepare-and-measure QKD system has not been established well. Therefore, exploring and evaluating practical security in high-speed QKD systems is currently important and meaningful.

In this paper, we establish a security analysis model for a QKD system under a class of non-intercept-resend attacks -- the orthogonal state attack that randomly transmits hacking pulse orthogonal to Alice's signal pulses to extract key information. Building upon this theoretical framework, we propose a muted attack that does not require intercept-resend operation, which exploits the vulnerability of a high-speed SPAD used in a QKD system. Unlike blinding attack~\cite{Lydersen2010,anqi2016,zhihao2020,gao2022}, the muted attack does not inject bright light to switch the SPAD to the linear mode, but instead sends multi-photon pulses to maintain its working in Geiger mode continuously. Only when Alice's state is orthogonal to that sent by Eve, Bob's SPAD is capable of responding normally to the photons sent by Alice. Otherwise, Bob's SPAD is inactive. Combined with Bob's basis selection information, Eve is able to learn nearly all of the key information. We then specifically propose a muted attack belonging to the class of orthogonal state attack. The feasibility of muted attack is verified by experimental tests on a SPAD equipped with a width discriminator, operating at the gate frequency \SI{1}{\giga\hertz}. Furthermore, we analyze the implementation security impact of the orthogonal state attack based on the theoretical model and experimental data of muted attack and dead-time attack under the condition that Alice and Bob are unaware of the attack.


\section*{Results}\label{sec2}

\subsection*{Security analysis model under orthogonal state attack}
\label{model}
In this section, we build the security analysis model of the BB84 QKD system considering the detection response under orthogonal state attack when Alice and Bob are unaware of Eve's presence. The BB84 QKD protocol uses four distinct quantum states of a single photon to encode key information. Taking a polarization-encoded QKD system as an example, horizontal ($\ket{H}$) and vertical ($\ket{V}$) polarizations constitute the $Z$ basis, while diagonal ($\ket{D}$) and anti-diagonal ($\ket{A}$) polarizations constitute the $X$ basis. Alice randomly transmits one of the four quantum states through the quantum channel, and Bob randomly chooses the $Z$ basis or the $X$ basis for the projection measurement of the quantum states received. Subsequently, Alice and Bob compare their basis selection information over the classical channel and retain only the secret key corresponding to matching bases.

\begin{figure*}[htbp]
\begin{center}
    \includegraphics[width=13cm]{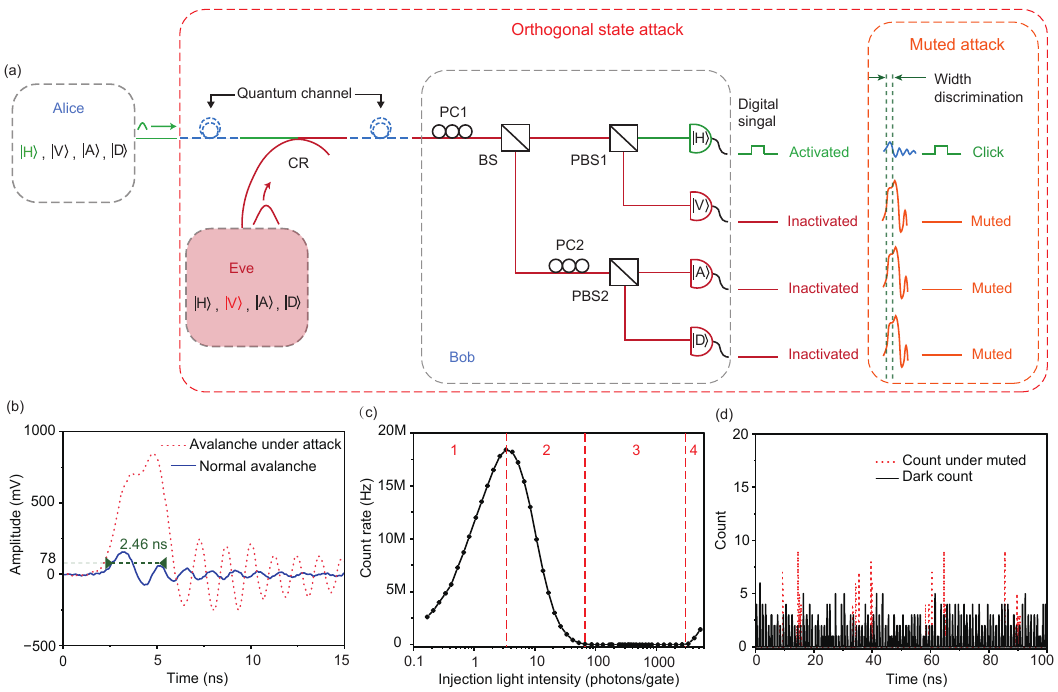}
\caption {Orthogonal state attack framework and outline of muted attack. (a) Schematic of an orthogonal state attack on the receiver of a passive-basis-selection BB84 QKD system (red dashed box). The SPADs receiving the hacking pulses and their digital outputs indicate whether the detectors are activated for photons sent by Alice. In the schematic of the muted attack (orange dashed box), the avalanche signals and their corresponding digital outputs indicate the click events associated to with detection of particular quantum states.
CR, coupler; PC, polarization controller; BS, 50:50 beam splitter; PBS, polarizing beam splitter.
(b) Avalanche current under two scenarios and the operating logic of the width discriminator. The red dashed line represents the avalanche current of the SPAD under the muted attack, while the blue solid line indicates the avalanche current induced by a single photon in the absence of an attack. The green dashed line represents the amplitude threshold (\SI{78}{\milli\volt}) and the width threshold (\SI{2.46}{\nano\second}) of the width discriminator.
(c) The count rate of the SPAD under the muted attack at different light intensities.
(d) The original dark counts and the residual counts under the muted attack.
The click statistics of the SPAD were collected over a 10-second duration under muted attacked and non-attacked conditions. The solid line represents the dark counts recorded with no attack, while the dotted line represents the counts when the SPAD receives 300 photons per gate. The variation in the relative timing between the end of the dead time and the arrival of the hacking pulse results in a bimodal distribution of residual counts.
}
	\label{fig:main}
\end{center}
\end{figure*}

The orthogonal state attack is feasible for both active and passive basis-selection QKD systems. Here, taking the passive basis-selection BB84 QKD protocol as an example, the strategy of the orthogonal state attack is shown in the~\cref{fig:main}(a). With no interception of Alice's sending state, 
Eve generates a multi-photon hacking pulse that is randomly modulated into one of the four quantum states.
When reaching Bob and passing through the beam splitter (BS), the hacking pulse is evenly divided into two beams. On the port where Bob matches the same basis as Eve, the photons reach a single detector. On the other port, the photons are split evenly between two detectors. However, these three detectors cannot register any clicks in the target bit time window due to certain security flaws, thus failing to respond to the incoming photon properly. The fourth detector, which does not receive the hacking pulse, instead, normally detects the signal pulse sent by Alice. Once Bob’s detector receives the signal pulse and registers a click event, he announces the basis information through a classical channel. Based on the basis information, Eve can infer that the quantum state Bob registered is orthogonal to that she sent on the same basis. Therefore, in the ideal scenario, this hacking strategy is sufficient for her to obtain the entire sifted key.

Here, we discuss within the scenario of a BB84 QKD protocol employing two decoy states. In addition to using the signal state with the mean photon number $\mu$ to transmit key information between Alice and Bob, Alice also emits two decoy states characterized by different mean photon numbers, $\nu_1$ and $\nu_2$. These mean photon numbers satisfy $\mu > \nu_1 > \nu_2$. 
Usually, the $Z$ basis is used to generate the key and the $X$ basis is used for parameter estimation. The lower bound of the key rate can be estimated as~\cite{2004}
\begin{equation}
R\geq q\{-Q_\mu f(E_\mu)H_2 (E_\mu)+Q_1[1-H_2(e_1)]\},
\label{equ.rate}
\end{equation}
where $H_2(x) = -x \log_2(x) - (1-x) \log_2(1-x)$ is the binary Shannon entropy function. 
The value of $q$ is implementation-dependent. $q=\frac{1}{2}$ for the BB84 protocol, since Alice and Bob select different bases approximately half of the time, and if one uses the efficient BB84 protocol~\cite{Lo2005.2},  $q\approx1$.
$Q_\mu$ denotes the overall gain of Bob's detector. The parameter $f$ represents the efficiency of error correction and $E_\mu$ represents the overall quantum bit error rate (QBER). $Q_1$ represents the lower bound for the single-photon gain and $e_1$ denotes the upper bound for the single-photon phase error rate.


Considering the orthogonal state attack, a general theoretical model can be established to describe the impact of this attack on the detection probabilities of the detectors. Using this model, one can further analyze the influence of the orthogonal state attack on the key rate of QKD systems.
Once Eve successfully implements the orthogonal state attack, Bob's valid clicks ideally depend on whether Eve sends a quantum state that is orthogonal to the signal state sent by Alice. In a more practical case, the hacking pulse may only have a limited probability of controlling a response from the detector, or the detector may still respond weakly under attack. These factors influence the overall gain of Bob’s detector, which can be represented as 
\begin{equation}
Q_{\mu}=q^{(=)}Q_{EA}^{=}+q^{(\perp)}Q_{EA}^{\perp}+q^{(\times)}Q_{EA}^{\times}.
    \label{eq.gain}
\end{equation}
Here, $Q_{EA}$ represents Bob's gain under the orthogonal state attack. The superscript symbol indicates three distinct scenarios of Eve's hacking pulses. ($=$) represents the scenario that the quantum state of the hacking pulse is the same as the pulse emitted by Alice; ($\perp$) is the scenario that the quantum state of the hacking pulse is orthogonal to the pulse emitted by Alice; ($\times$) indicates the scenario that the hacking pulse is modulated on a different basis from Alice's. $q$ defines the probability weight for each scenario in practical implementations.
In the case where the legitimate parties are unaware of the hacker's presence, 
Bob cannot determine the scenario in which the events occur.



For each injected hacking pulse, Eve expects the target detector to produce a valid response and contribute to the gain, denoted $Q_{target}$. The other three detectors are rendered inactive upon receiving the hacking pulses, with probability $P_{attack}$.
Accordingly, the overall gain $Q_{\mu}$ can be expressed as
\begin{equation}
     Q_{\mu} = \sum_{x\in\{^{=},^{\perp},^{\times}\}} q^{(x)} \times Q_{target} \times P_{attack}.
\end{equation}
For the target detector that is not suppressed by the attack, the detection probability is mainly determined by normal detection events induced by signal photons $Q_A$ and dark count $Y_0$. In certain attack scenarios, the target detector may be successfully controlled with a probability $P_{a,t}^{(x)}$. Nevertheless, even under such conditions, it may still generate clicks with a residual probability $P_{click,t}^{(x)}$.
Therefore, $Q_{target}$ can be calculated as
\begin{equation}
    Q_{target}=(1-P_{a,t}^{(x)})(Q_A+Y_0)+P_{a,t}^{(x)}P_{click,t}^{(x)}.
\end{equation}
For the three attacked SPADs, they are typically rendered inactive with a high probability $P_{a,i}^{(x)}$. However, in certain attack scenarios, the attacked SPADs may still produce clicks with probability $P_{click,i}^{(x)}$. Therefore, $P_{attack}$ can be expressed as
\begin{equation}
    P_{attack}=\prod_{i=1}^{3} P_{a,i}^{(x)}(1-P_{click,i}^{(x)}) + (1-P_{a,i}^{(x)})(1-Y_0).
\end{equation}

In the orthogonal state attack, the probability of erroneous click events may differ depending on the distinct quantum states of the hacking pulses modulated by Eve, and the overall QBER can therefore be calculated as
\begin{equation}
    E_\mu Q_\mu=e_{EA}Y + e_{det}Q_A,
    \label{}
\end{equation}
where $e_{det}$ is the probability that a photon hits the erroneous detector, and $Q_A$ represents the gain that the target detector responds to the signal pulse. 
We decompose the probability of erroneous clicks into two main contributions, the probability that the SPAD orthogonal to the signal state produces a response $P_{ort}$, and the probability that the other three detectors do not click $P_{nort}$.
The probability of erroneous click events $e_{EA}Y$ can be calculated as
\begin{equation}
     e_{EA}Y=\sum_{x\in\{^{=},^{\perp},^{\times}\}} q^{(x)} \times P_{ort} \times P_{nort}.
\end{equation}
For the detector orthogonal to the signal state, when no hacking pulse is applied, its response probability is mainly determined by dark counts $Y_0$. When it is exposed to a hacking pulse, it is inactivated with a probability $P_{a,e}^{(x)}$. In certain attack scenarios, the attacked SPAD may still produce clicks with a probability $P_{click,e}^{(x)}$. Therefore, $P_{ort}$ can be expressed as
\begin{equation}
    P_{ort}=(1-P_{a,e}^{(x)})Y_0+P_{a,e}^{(x)}P_{click,e}^{(x)}.
\end{equation}
For the other three SPADs, they are inactivated with a probability $P_{a,i'}^{(x)}$ under attack, while still producing clicks with a probability $P_{click,i'}^{(x)}$. Therefore, $P_{nort}$ can be expressed as
\begin{equation}
    P_{nort}=\prod_{i'=1}^{3} [P_{a,i'}^{(x)}(1-P_{click,i'}^{(x)}) + (1-P_{a,i'}^{(x)})(1-Y_0)].
\end{equation}


\subsection*{The working principle of muted attack}


Following the framework of orthogonal state attack, this idea is applicable to a high-speed QKD system equipped with specialized sinusoidal gating SPADs that employ the low-pass filter (LPF) and the width discriminator~\cite{HE2017,he2023}. The implementation of this high-speed SPAD introduces a potential vulnerability that can be exploited by an eavesdropper, Eve, to compromise the security of the high-speed decoy-state BB84 QKD system. We call this specific type of orthogonal state attack as the muted attack.

The muted attack is applicable to QKD systems with both active and passive basis selection. Taking a passive-basis-selection polarization-encoding QKD system as an example, we assume that Eve can acquire the polarization reference of the legitimate parties by monitoring the system's calibration process, analyzing the real-time QBER feedback, or employing a Trojan-horse attack.
Then Eve injects optical pulses with the sufficient number of photons into Bob, each of which is randomly modulated to be one of the four quantum states as shown in \cref{fig:main}(a). Three SPADs in the measurement apparatus receive the hacking pulse. It is worth noticing that the multiphoton hacking pulse can reach Bob's detector in the detection time window, unlike the dead time attack in which the hacking pulse arrives outside the time window. Upon receiving the hacking pulse, the SPAD generates a wide avalanche pulse, which is identified and filtered out by the width discriminator, thus no clicks registered. As a result, each hacking pulse sent by Eve causes Bob’s three SPADs out of four to become muted. Even if the Alice's pulse and the hacking pulse arrive simultaneously at a certain SPAD, the muted one does not produce any click. Only when Alice's pulse reaches the unique non-muted SPAD could it produce a click as normal.

The key point for Eve to successfully launch the muted attack is to keep the SPAD constantly generating wide avalanche by injecting multi-photon hacking pulses and thus being muted. In the SPAD equipped with a LPF, a common issue arises where consecutive avalanche events may merge into a wide avalanche signal, leading to an increased after-pulse probability. To suppress this effect, a width discriminator is employed to filter the wide avalanche signals. when the signal amplitude reaches the threshold, the discriminator selectively passes only the pulses that fall within the discrimination width window. Any pulse exceeding the width of the discrimination is rejected by the discriminator logic circuit, and thus no click event is registered~\cite{HE2017,he2023}. However, the width discriminator introduces a loophole that allows Eve to eavesdrop the secret key. When the SPAD receives the hacking pulse, a wide avalanche is formed as red dotted curve in~\cref{fig:main}(b), which is rejected by the width discriminator to register as a valid click. That is, the SPAD is muted under injection of the multi-photon hacking pulse.
Consequently, the muted attack specifically targets SPADs equipped with width discrimination. It is noted that the muted attack is also applicable to phase-encoding QKD implementation but not fully applicable to time-bin-phase encoding one.

\subsection*{Experimental demonstration}
We conduct the experimental verification of the muted attack on a 1-\si{GHz} SPAD equipped with a variable width discriminator~\cite{HE2017,he2023}, which can be used in the receiver unit of high-speed QKD systems for single photon detection~\cite{Wang2018}.
Once the width of each muted avalanche exceeds the discrimination width (\SI{2.46}{\nano\second}) at the discriminator threshold (\SI{78}{\milli\volt}) as shown in~\cref{fig:main}(b), the SPAD under test does not register a click but still enters a 23-\si{ns} dead time. The core operation parameters of the SPAD under test are shown in~\cref{tab:parameters}. 

To reliably induce a wide avalanche instead of a normal one, the width of the hacking pulse must span two consecutive gating windows, so that the avalanche signals from adjacent gates merge into a single extended response. Therefore, the hacking pulse width is chosen to be \SI{2}{\nano\second}.
Due to the pronounced current tail following a wide avalanche, it is necessary to suppress this detrimental effect to maintain the ``muted” state of the SPAD. 
Thus, the relaxation period of the hacking pulses is best matched to the dead time of \SI{23}{\nano\second} so that there is enough time for carrier relaxation and recovery. Once the dead time ends, the SPAD can be continuously controlled again by the hacking pulse.
To demonstrate the muted attack, the optical hacking pulse applied in the test features the wavelength of \SI{1550}{\nano\meter}, the repetition rate of \SI{40}{\mega\hertz}, and the pulse width of \SI{2}{\nano\second}. 
The mean photon number of each injected hacking pulse ranges from 0.1 to 5000, 
with steps corresponding to $1~\si{dBm}$ power increments.
  \begin{center}
  \captionof{table}{The parameters of the SPAD under test}
    \begin{tabular}{lc}
      \hline
  	\textbf{Parameter} & \textbf{Value} \\
  	\hline
  	Gating frequency & \SI{1 }{\giga\hertz} \\
  	Dead time & \SI{23 }{\nano\second} \\
  	Detection efficiency & $20.6~\%$ \\
       Dark count rate & $1.5\times10^{-7}$/\text{gate} \\
 	\hline
    \end{tabular}
    \label{tab:parameters}
  \end{center}



  		

The SPAD's counting rate under the muted attack is recorded as shown in \cref{fig:main}(c). The trend of the counting rate is divided into four regions. 
Region 1 is the rising stage, where the count rate of the SPAD enhances with increasing number of photons reaching the APD. Region 2 is the declining stage, in which the increasing number of incident photons leads to more periods being muted, resulting in the continuous decrease in count rate of photons. Region 3 is the muted stage, which is the work area of the attack. Although the number of injected photons increases, there is nearly no click registered. That is, the hacking pulse almost 100\% controls the SPAD to turn into a muted status $P_{a,i}\approx1$. The discriminator enters dead time after detecting the wide avalanche signal. Once the dead time ends, the SPAD receives the hacking pulse again, being muted once more. This cycle of muted detection is periodically repeated.
In Region 4, as the photon count continues to increase and reaches 3000 photons per gate or more, the SPAD's count rate rises back.
This is because the increased avalanche charge leads to stronger tail pulses, which ultimately reaches the discrimination threshold.
Eve gradually loses control over the SPAD.

When Eve implements the muted attack on an actual QKD system, she tends to minimize her own cost by injecting as few photons as possible. 
Eve aims to control the SPADs response while minimizing the required injected optical power. Increasing the photon number can further suppress the click probability of the SPAD, however, it also increases the probability of photon leakage due to the finite extinction ratio of the polarizing beam splitter (PBS) in the QKD receiver, thereby introducing additional QBER.
When the number of photons injected by Eve gradually increases to 150 photons, the count rate decreases to \SI{134}{\hertz} ($P^{\times}_{click,t}=1.34\times10^{-7}$), approaching the dark count level \SI{150}{\hertz}. At this point, even if the SPAD clicks due to a hacking pulse, it would be interpreted as a dark count of the SPAD and does not raise Bob's QBER. Thus, it is essential to ensure that each SPAD receives at least 150 photons. That is, when Bob selects the same basis as Eve, there are 300 photons reaching a single SPAD, at which point the counting rate of this SPAD drops to \SI{32}{\hertz} ($P^=_{click,t}=3.2\times10^{-8}$).

The count rate of the muted SPAD receiving 300 photons per gate is lower than the dark count level, but the time distribution of its click events shows a significant difference from the dark count. As shown in~\cref{fig:main}(d), the clicks of the muted SPAD are not randomly distributed like the dark count but are concentrated to exhibit statistical characteristics of two distinct peaks within each period. The appearance of the first peak is due to the count caused by the afterpulse once the dead time ends. The second peak indicates that when the wide avalanche is formed, the SPAD is still in the dead time. After the dead time ends, the tail pulses shown in \cref{fig:main}(b) may satisfy the conditions for the output of the discriminator, triggering a click.

\subsection*{Impact on the key rate of QKD system}

In this study, we present a comprehensive security analysis of the orthogonal state attacks, including the muted attack and the dead time attack. To analyze the impact of orthogonal state attack on the key distributed between Alice and Bob, the key rate generated under orthogonal state attack is simulated based on the proposed theoretical model and the corresponding experimental data. It is assumed that Alice and Bob are unaware of Eve's presence.



The theoretical simulation is based on the decoy-state BB84 QKD protocol~\cite{wang2005,lo2005,youpian,youpian2}, which is the scheme most commonly implemented in current QKD deployments. In the scenario of the muted attack, the hacking pulse and the signal pulse arrive at the detector simultaneously, and the number of photons in the hacking pulse is sufficient to ensure that the detector forms the wide avalanche rather than a normal avalanche. Therefore, each hacking pulse almost completely mutes three detectors, implying that $P_{a,i}\approx 1$. However, due to the residual counts in the muted attack, the detector under attack may still register a click event. Thus, the probability of a click at the target gate, $P_{click,i}$, is non-zero.

In the ideal case, when Eve's hacking pulse causes three SPADs to be muted, the other one is not affected by the hacking pulse, thereby ensuring that the QBER does not increase. When Bob's SPAD is muted, any detection event registered is attributed to the dark counts of the SPAD. According to the security model proposed in Sec. Security analysis model under orthogonal state attack, the key rate under the ideal muted attack is obtained as indicated by the red dotted line in~\cref{fig:key}. It is noted that when Eve conducts the muted attack by injecting hacking pulses, the probability that Bob successfully receives the signal-state photon sent from Alice is $25\%$. This probability is halved compared to $50\%$ of correct basis selection in the absence of the attack. The key rate with attack thus is half that without attack (black solid line) in short transmission distance.

  \begin{center}
    \includegraphics[width=8cm]{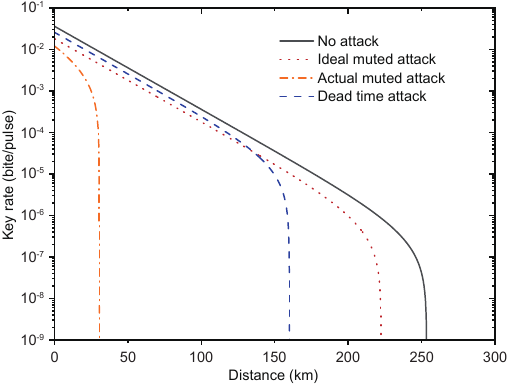}
    \captionsetup{width=\linewidth}
    \captionof{figure}{The simulation of the key rate of the QKD system under different conditions. The black solid line represents the secure key rate of the QKD system in the absence of attacks. The other three curves correspond to the overestimated key rates under different orthogonal state attack scenarios, where the legitimate users are unaware of Eve’s presence, but all the key information shared between Alice and Bob is in fact insecure and has been obtained by Eve. The red dotted line denotes the ideal muted attack scenario, the orange dash-dotted line represents the practical muted attack scenario, and the blue dashed line corresponds to the dead time attack scenario.}
    \label{fig:key}
  \end{center}


In the practical case, due to the limited extinction ratio of the PBS, there is a probability that the hacking pulse leaks photons to the non-muted SPAD when Bob and Eve choose the same basis, which may increase the QBER.
In our test, the extinction ratio of two cascaded PBSs is approximately $43~\si{\decibel}$, which is a typical value in a QKD system. Consequently, the key rate curve under this actual attack condition is represented by the orange dash-dotted line in~\cref{fig:key}, which is significantly lower compared to the ideal attack scenario. The details of the key rate estimate are presented in section Methods. It is worth noticing that the secret key under the muted attack calculated by Alice and Bob is indeed overestimated, since they are unaware of Eve's attack. In fact, Eve obtains almost all the keys via the muted attack and the actual secure key would be nearly zero.



The theoretical model under the orthogonal state attack is also suitable for analyzing the impact of dead time attack on the practical security of a QKD system. According to the dead time attack presented in Ref.~\cite{deadtimeattack2011}, it is assumed that the hacking pulses arrive at the detector at time $t_{B,i}$, shortly (less than dead time $\tau_D$) before the signal pulse reaches time $t_i$ and is still outside Bob's time window ($\frac{\Delta_{tw}}{2} < t_i - t_{B,i} < \tau_D$). The mean photon number of the hacking pulse is $\mu=16$, which is consistent with the experimental data provided in~Ref.~\cite{deadtimeattack2011}, to maximize the mutual information between Eve and Bob.

Due to the fact that the number of photons in the hacking pulse consists only of a dozen photons per pulse, the probability of the detector response to the hacking pulse $P_{a,i}=1-e^{-n\eta}<1$. Here, $n$ represents the mean photon number of the hacking pulse, and $\eta$ is the detection efficiency of the detector.
When the detector receives a hacking pulse and enters its dead time, no count is recorded for the detector under attack, $P_{click,i}=0$.
Based on the decoy-state BB84 QKD protocol, we simulate the overestimated key rate between Alice and Bob, as shown in the blue dashed line in~\cref{fig:key}. Since the count rate drops to zero once the detector under attack enters the dead time, the overall dark count rate appears to be reduced from Bob’s perspective. In the dead time attack, the target detector does not produce residual counts, resulting in a lower QBER and a higher key rate estimated by Alice and Bob at short transmission distances compared to the ideal muted attack scenario.
However, a lower but fake dark count rate under the muted attack allows Alice and Bob to still estimate the positive key rate at a longer maximum transmission distance than that of the dead time attack.

\section*{Discussion}\label{sec12}
Compared with the well known blinding attack in QKD systems, the orthogonal state attack represents a fundamentally different detector control strategy and offers several practical advantages. 
First, the orthogonal state attack does not need the intercept-resend operation but manipulates the detector responses by injecting optical pulses with precisely controlled timing. 
Consequently, the implementation of the orthogonal state attack does not critically
rely on advanced components such as high-speed detectors, high-speed modulators, or narrow-pulse high-power laser sources.
The orthogonal state attack significantly reduces the operational complexity for Eve and makes it more suitable for high-speed QKD systems.
Furthermore, unlike blinding attacks, the orthogonal state attack does not require switching the SPAD into the linear mode by injecting strong light but operates at relatively low optical power levels. As a result, countermeasures based on the detection of strong light injection could become ineffective.

The orthogonal state attack also exhibits several inherent limitations compared with blinding attacks. Although both attacks can, in principle, allow Eve to learn all the sifted key, the orthogonal state attack introduces a more reduction in the sifted key between Alice and Bob (typically to about 25\%), whereas blinding attacks can retain up to 50\% in active-basis-selection QKD systems and nearly 100\% in passive-basis-selection implementations. In addition, the orthogonal state attack cannot deterministically induce clicks, but the trigger pulse used in blinding attack can manipulate the detection efficiency up to 100\%. Within this framework, the muted attack may introduce additional QBER due to the finite extinction ratio of the PBS. Fortunately, this effect can be mitigated by timing the hacking pulses to the avalanche transition region~\cite{kang2026}.

Orthogonal state attacks adopt a low-photon-number injection approach and lead to the near absence of system parameter variations, which is challenging to detect by communicating parties. Therefore, establishing effective active countermeasures for high-speed QKD systems is important and necessary. An effective countermeasure against dead-time attack is known--whenever any detector registers an optical pulse, all detectors should be forced into dead time simultaneously. Nevertheless, this countermeasure fails against the muted attack, in which Eve precisely injects her pulse inside the detection time window to control the detector. Fortunately, monitoring the filtering behavior of the width discriminator is a direct and useful method. If the width discriminator shows a frequently active filtering phenomenon for wide avalanche pulses, it can be inferred that Eve may be conducting an attack, and her attack intention can be exposed timely. Furthermore, by analyzing the time distribution characteristics of the dark count of Bob's SPAD, the periodic and regular distribution (as shown in~\cref{fig:main}d) indicates that Eve might have controlled Bob's detection events through the muted attack during the key distribution process.



In high-speed QKD systems, orthogonal state attacks do not require intercept-resend operations, making them unaffected by narrow time windows, which significantly impacts the security performance. 
This paper establishes a general theoretical analysis model for the orthogonal state attack with the aim of analyzing the potential impact of such attacks on the implementation security of the QKD system. 
Based on this model, we propose the muted attack by exploiting the security loophole introduced by the width discriminator in SPAD. 

Unlike traditional attacks, in this muted attack, Eve neither needs to conduct intercept-resend operations via the quantum channel, nor does she need to inject strong light to switch the SPAD into linear mode.
More importantly, Eve can obtain nearly all the keys between Alice and Bob via this muted attack.
Experiment results show that, by applying a few hundreds of photons, Eve can achieve the purpose of eavesdropping by merely matching the period of the hacking pulse with the dead time of the SPAD.

The results of the implementation security analysis show that orthogonal-state attacks allow key establishment
when legitimate parties are unaware of the presence of Eve, compromising the security of QKD.
This study highlights the threat of non-intercept-resend attacks to the practical security of high-speed QKD systems, as well as the potential security issues introduced by advanced devices and technologies in these systems, which deserve further deep research and more attention.
We believe that this research is timely and will drive further development in the security of practical high-speed QKD systems.

\section*{Methods}
\label{method}

\subsection*{Overestimated key rate under the muted attack}

We used the theoretical model of the orthogonal state attack to calculate the overestimated key rate of Alice and Bob under the muted attack.
Due to the muted attack, Eve determines Bob's selection of basis for each time, which means Bob loses control over $q$ as shown in~\cref{equ.rate}.
In the ideal scenario, when Bob's SPAD is muted, any detection event registered during this period is attributed to the dark counts of the SPAD.
One SPAD receives 300 photons per gate, the click probability is $3.2 \times 10^{-8}/\text{gate}$, and two SPADs each receive 150 photons per gate, the click probability is $1.34 \times 10^{-7}/\text{gate}$. 
The other SPAD is not affected by the hacking pulse.
Here, we denote the residual count probability by $Y_{300} = 3.2 \times 10^{-8}$, $Y_{150} = 1.34 \times 10^{-7}$. The actual dark count probability is $Y_0=1.5 \times 10^{-7}/\text{gate}$.

For a SPAD, the response probability to the signal pulses sent by Alice is 
\begin{equation}
    Q_A=\sum_{i=0}^{\infty} \frac{\lambda^i e^{-\lambda}}{i!}  P_{\text{det}}(i)=1-e^{-\eta\lambda},
    \label{equ.3}
\end{equation}
where $\lambda =\mu,\nu_1,\nu_2$, and $\eta$ denote the overall transmission and detection efficiency between Alice and Bob.
Assume that Alice prepares four quantum states with equal probability. Therefore, when Eve performs the muted attack, regardless of the quantum state Alice emits, she has $q=25\%$ probability of correctly guessing and sending the correct attack state.
Next, we analyze the gain of Bob's SPAD in the context of the three scenarios.
When Eve emits an hacking pulse with the quantum state identical to Alice's, 
\begin{equation}
\begin{split}
Q_{EA}^{= } &= \frac{1}{4} \times [(1-P_{2n})(Q_A+Y_0)+P_{2n}Y_{300}] \\
&\quad \times [(0)+(1)(1-Y_0)] \\
&\quad \times [P_{n}(1-Y_{150})+(1-P_{n})(1-Y_0)] \\
&\quad \times [P_{n}(1-Y_{150})+(1-P_{n})(1-Y_0)].
\end{split}
\end{equation}
Due to the arrival of $2n=300$ photons from the hacking pulse in the target detector, the probability of the SPAD being muted is $P_{2n}\approx 1$. The reason for registering a click is the residual count under the attack. The probability of the other SPAD on the same base being muted is zero, and the reason for it to register a click is the dark count of the SPAD itself.
When Eve sends a quantum state that is orthogonal to Alice's on the same basis,
\begin{equation}
\begin{split}
Q_{EA}^{\perp} &= \frac{1}{4}\times (1)(Q_A+Y_0)\\
&\quad \times [P_{2n}(1-Y_{300})+(1-P_{2n})(1-Y_0)]\\
&\quad \times [P_{n}(1-Y_{150})+(1-P_{n})(1-Y_0)]\\
&\quad \times [P_{n}(1-Y_{150})+(1-P_{n})(1-Y_0)].
\end{split}
\end{equation}
Since Eve sends the correct quantum state, the probability that the target detector is not muted is 1, which allows it to properly receive the signal pulse.
When Eve sends a quantum state on a basis different from Alice's,
\begin{equation}
\begin{split}
Q_{EA}^{\times} &= \frac{1}{2}[(1-P_{n})(Q_A+Y_0)+P_{n}Y_{150}]\\
&\quad \times[P_{n}(1-Y_{150})+(1-P_{n})(1-Y_0)]\\
&\quad \times[P_{2n}(1-Y_{300})+(1-P_{2n})(1-Y_0)]\\
&\quad \times [(0)+(1)(1-Y_0)].
\end{split}
\end{equation}
Therefore, Bob's overall gain is
\begin{equation}
    Q_{\mu}=Q_{EA}^{=}+Q_{EA}^{\perp}+Q_{EA}^{\times}.
    \label{eq.5}
\end{equation}

The probability of Bob's SPADs giving an erroneous response during Eve's implementation of a muted attack is as follows.
When Eve sends a hacking pulse identical to Alice's quantum state, the probability of Bob experiencing bit error is
\begin{equation}
\begin{split}
e_{EA}Y^= &=\frac{1}{4}[(1)Y_0+(0)]\\
&\quad \times [P_{2n}(1-Y_{300})+(1-P_{2n})(1-Q_A)(1-Y_0)]\\
&\quad \times[P_{n}(1-Y_{150})+(1-P_{n})(1-Y_0)]\\
&\quad \times[P_{n}(1-Y_{150})+(1-P_{n})(1-Y_0)].
\end{split}
\end{equation}
Due to Eve choosing the wrong quantum state, the target detector is attacked, and the clicks from the other detector within the same basis are the main cause of bit error. The probability of the other detector being attacked is zero, and the registered click events arise from its dark counts.
When Eve sends a quantum state that is orthogonal to Alice's on the same basis,
\begin{equation}
\begin{split}
e_{EA}Y^{\perp} &=\frac{1}{4}[(1-P_{2n})Y_0+P_{2n}Y_{300}]\\
&\quad \times [(0)+(1)(1-Q_A)(1-Y_0)]\\
&\quad \times [P_{n}(1-Y_{150})+(1-P_{n})(1-Y_0)]\\
&\quad \times [P_{n}(1-Y_{150})+(1-P_{n})(1-Y_0)]
\end{split}
\end{equation}
When Eve sends a quantum state on a basis different from Alice's,
\begin{equation}
\begin{split}
e_{EA}Y^{\times} &=\frac{1}{2}[P_{n}(1-Y_{150})+(1-P_{n})(1-Q_A)(1-Y_0)]\\
&\quad \times [(1-P_{n})Y_0+P_{n}Y_{150}]\\
&\quad \times [P_{2n}(1-Y_{300})+(1-P_{2n})(1-Y_0)]\\
&\quad \times [(0)+(1)(1-Y_0)].
\end{split}
\end{equation}
The overall QBER is
\begin{equation}
    E_\mu Q_\mu=e_{EA}Y^= + e_{EA}Y^{\perp} + e_{EA}Y^{\times} +e_{det}Q_A,
    \label{}
\end{equation}
where $e_{det}$ is the probability that a photon hits the erroneous detector. 

Here we employ the analytical approach proposed by~Ref.~\cite{youpian} to evaluate the lower bound of $Y_1$ and the upper bound of $e_1$. Based on~\cref{equ.rate}, the key rate under the ideal muted attack can be obtained as indicated by the red dotted line in~\cref{fig:key}.

In the practical scenario, we calculate the bit rate with the extinction ratio of PBS being $43~\si{\decibel}$, when Bob's $\ket{V}$ detector receives 300 photons, there is $1.5\%$ probability that one photon is leaked to the $\ket{H}$ detector, denoted $n_L=0.015$.
If Eve guesses wrongly about the quantum state, the leaked photons increase the QBER.
The residual response probability of the non-muted detector attributed to the attack, is given by
\begin{equation}
    Q_E=\sum_{n=0}^{\infty} \frac{n_L^n e^{-n_L}}{n!}  P_{\text{det}}(n)=1-e^{-n_L\eta_D}.
    \label{equ.4}
\end{equation}

In the case of an actual attack, the gain $Q_{\mu}^{'}$ of Bob's SPAD can be calculated as
\begin{equation}
\begin{split}
Q_{EA}^{='} &= \frac{1}{4}[(1-P_{2n})(Q_A+Y_0)+P_{2n}Y_{300}]\\
&\quad \times [(0)+(1)(1-Y_0)(1-Q_E)]\\
&\quad \times [P_{n}(1-Y_{150})+(1-P_{n})(1-Y_0)]\\
&\quad \times [P_{n}(1-Y_{150})+(1-P_{n})(1-Y_0)],
\end{split}
\end{equation}
\begin{equation}
\begin{split}
Q_{EA}^{\perp'} &= \frac{1}{4}(1)(Q_A+Q_E+Y_0)\\
&\quad \times[P_{2n}(1-Y_{300})+(1-P_{2n})(1-Y_0)]\\
&\quad \times[P_{n}(1-Y_{150})+(1-P_{n})(1-Y_0)]\\
&\quad \times[P_{n}(1-Y_{150})+(1-P_{n})(1-Y_0)],
\end{split}
\end{equation}
\begin{equation}
\begin{split}
Q_{EA}^{\times'} &= \frac{1}{2}[(1-P_{n})(Q_A+Y_0)+P_{n}Y_{150}]\\
&\quad \times [P_{n}(1-Y_{150})+(1-P_{n})(1-Y_0)]\\
&\quad \times [P_{2n}(1-Y_{300})+(1-P_{2n})(1-Y_0)]\\
&\quad \times [(0)+(1)(1-Y_0)(1-Q_E)],
\end{split}
\end{equation}
\begin{equation}
    Q_{\mu}^{'}=Q_{EA}^{='}+Q_{EA}^{\perp'}+Q_{EA}^{\times'}.
    \label{eq.5}
\end{equation}

In the case of an actual attack, the over all QBER can be calculated as
\begin{equation}
\begin{split}
e_{EA}Y^{='} &=\frac{1}{4}[(1)(Q_E+Y_0)+(0)]\\
&\quad \times [P_{2n}(1-Y_{300})+(1-P_{2n})(1-Q_A)(1-Y_0)]\\
&\quad \times[P_{n}(1-Y_{150})+(1-P_{n})(1-Y_0)]\\
&\quad \times[P_{n}(1-Y_{150})+(1-P_{n})(1-Y_0)],
\end{split}
\end{equation}
\begin{equation}
\begin{split}
e_{EA}Y^{\perp'} &=\frac{1}{4}[(1-P_{2n})Y_0+P_{2n}Y_{300}]\\
&\quad \times [(0)+(1)(1-Q_E)(1-Q_A)(1-Y_0)]\\
&\quad \times [P_{n}(1-Y_{150})+(1-P_{n})(1-Y_0)]\\
&\quad \times [P_{n}(1-Y_{150})+(1-P_{n})(1-Y_0)],
\end{split}
\end{equation}
\begin{equation}
\begin{split}
e_{EA}Y^{\times'} &=\frac{1}{2}[P_{n}(1-Y_{150})+(1-P_{n})(1-Q_A)(1-Y_0)]\\
&\quad \times [(1-P_{n})Y_0+P_{n}Y_{150}]\\
&\quad \times [P_{2n}(1-Y_{300})+(1-P_{2n})(1-Y_0)]\\
&\quad \times [(0)+(1)(1-Q_E)(1-Y_0)],
\end{split}
\end{equation}
\begin{equation}
    E_\mu^{'} Q_\mu^{'}=e_{EA}Y^{='} + e_{EA}Y^{\perp'} + e_{EA}Y^{\times'} +e_{det}Q_A.
    \label{}
\end{equation}
The key rate curve under actual attack conditions is represented by the orange dash-dotted line in~\cref{fig:key}.


\backmatter


\section*{Data availability}
The datasets generated and analyzed during the current study are not publicly available due to ongoing analyzes and the inclusion of unpublished follow-up work, but are available from the corresponding author on reasonable request.

\section*{Code availability}
The codes generated during the current study are not publicly available due to ongoing analyzes and the inclusion of unpublished follow-up work, but are available from the corresponding author on reasonable request.

\section*{Acknowledgements}
This work was funded by the Quantum Science and Technology--National Science and Technology Major Project (2021ZD03007), National Natural Science Foundation of China (No.~62371459 and No.~62425507), and the Postdoctoral Fellowship Program of CPSF (GZC20252817).

\section*{Author contributions}
J.S. and J.C. conducted the experiment. D.H., S.W., and A.H. support the experimental conduct. J.S., Q.P., J.C., F.L., and A.H. analyzed the data. J.S., Q.P., J.L.,
and Z.C. compiled the code. J.S., Q.P., and A.H. wrote the paper with input from all authors. A.H. and S.W. supervised the project.

\section*{Competing interests}
The authors declare no competing financial or non-financial interests.


\section*{Figure legends}

\bmhead{Figure 1. Orthogonal state attack framework and outline of muted attack}

(a) Schematic of an orthogonal state attack on the receiver of a passive-basis-selection BB84 QKD system (red dashed box). The SPADs receiving the hacking pulses and their digital outputs indicate whether the detectors are activated for photons sent by Alice. In the schematic of the muted attack (orange dashed box), the avalanche signals and their corresponding digital outputs indicate the click events associated to with detection of particular quantum states.
CR, coupler; PC, polarization controller; BS, 50:50 beam splitter; PBS, polarizing beam splitter.

(b) Avalanche current under two scenarios and the operating logic of the width discriminator. The red dashed line represents the avalanche current of the SPAD under the muted attack, while the blue solid line indicates the avalanche current induced by a single photon in the absence of an attack. The green dashed line represents the amplitude threshold (\SI{78}{\milli\volt}) and the width threshold (\SI{2.46}{\nano\second}) of the width discriminator. 

(c) The count rate of the SPAD under the muted attack at different light intensities.

(d) The original dark counts and the residual counts under the muted attack.
The click statistics of the SPAD were collected over a 10-second duration under muted attacked and non-attacked conditions. The solid line represents the dark counts recorded with no attack, while the dotted line represents the counts when the SPAD receives 300 photons per gate. The variation in the relative timing between the end of the dead time and the arrival of the hacking pulse results in a bimodal distribution of residual counts.

\bmhead{Figure 2. The simulation of the key rate of the QKD system under different conditions}

The black solid line represents the secure key rate of the QKD system in the absence of attacks. The other three curves correspond to the overestimated key rates under different orthogonal state attack scenarios, where the legitimate users are unaware of Eve’s presence, but all the key information shared between Alice and Bob is in fact insecure and has been obtained by Eve. The red dotted line denotes the ideal muted attack scenario, the orange dash-dotted line represents the practical muted attack scenario, and the blue dashed line corresponds to the dead time attack scenario.

\end{document}